\documentclass[12pt]{iopart}
\usepackage[dvips]{graphicx}

\begin{document}

\title[Screening of classical Casimir forces by electrolytes]{Screening of 
classical Casimir forces by electrolytes in semi-infinite geometries}

\author{B. Jancovici\dag\ and L. {\v S}amaj\dag\ddag}
\address{\dag\ Laboratoire de Physique Th\'eorique, Universit\'e de
Paris-Sud, B\^atiment 210, 91405 Orsay Cedex, France
(Unit\'e Mixte de Recherche no. 8627 - CNRS)}
\address{\ddag\ Institute of Physics, Slovak Academy of Sciences,
D\'ubravsk\'a cesta 9, 845 11 Bratislava, Slovakia}

\eads{\mailto{Bernard.Jancovici@th.u-psud.fr}, \mailto{fyzimaes@savba.sk}}

\begin{abstract}
We study the electrostatic Casimir effect and related phenomena
in equilibrium statistical mechanics of classical (non-quantum)
charged fluids.
The prototype model consists of two identical dielectric slabs
in empty space (the pure Casimir effect) or in the presence
of an electrolyte between the slabs.
In the latter case, it is generally believed that the long-ranged
Casimir force due to thermal fluctuations in the slabs is screened
by the electrolyte into some residual short-ranged force.
The screening mechanism is based on a ``separation hypothesis'':
thermal fluctuations of the electrostatic field in the slabs can be
treated separately from the pure image effects of the ``inert'' slabs
on the electrolyte particles.
In this paper, by using a phenomenological approach under certain
conditions, the separation hypothesis is shown to be valid.
The phenomenology is tested on a microscopic model in which
the conducting slabs and the electrolyte are modelled by
symmetric Coulomb gases of point-like charges with different 
particle fugacities.
The model is solved in the high-temperature Debye-H\"uckel limit
(in two and three dimensions) and at the free fermion point of
the Thirring representation of the two-dimensional Coulomb gas.  
The Debye-H\"uckel theory of a Coulomb gas between dielectric walls
is also solved.

\end{abstract}


\medskip


\maketitle

\eqnobysec

\section{Introduction}
Two parallel metallic plates attract one another at zero temperature
due to fluctuations of the quantum electromagnetic field in vacuum.
This is the well-known Casimir effect; for a nice introduction see 
reference \cite{Duplantier}, for an exhausting review see \cite{Bordag}.
The extension of Casimir's result to arbitrary temperatures and to
general dielectric plates was made by Lifshitz et al. \cite{Lifshitz},
with a subsequent treatment of delicate points by Schwinger et al.
\cite{Schwinger}.
The Casimir force between two disconnected conductor objects of
different shapes, like a rectilinear wall and a sphere, is also 
treatable \cite{Duplantier} by using a method due to Derjaguin. 
The general form of the Casimir free energy for ideal-conductor walls 
of arbitrary smooth shapes was derived by Balian and Duplantier 
\cite{Balian1} using multiple scattering expansions
(for a review, see \cite{Balian2}). 
In the classical (or high-temperature) limit defined by the validity
of the equipartitioning energy law, the Casimir forces between
{\it disconnected} boundaries become purely entropic \cite{Feinberg},
and as such depend only on the geometry of the boundaries.
For accurate experimental measurements of the Casimir effect
see \cite{Mohideen}.

There is a permanent interest in the Casimir effect and related phenomena 
in equilibrium statistical mechanics of classical charged fluids, 
i.e. models of non-quantum charged particles which do not 
incorporate the magnetic part of the Lorentz force due to charge currents. 
Although such purely electrostatic models ignore the magnetic degree
of freedom (whose effect has the same magnitude as that of the
Coulomb force in the case of conducting walls), they can be used 
in more complex experimental situations to reveal microscopic mechanisms 
behind macroscopic electrostatics.
These classical electrostatic forces are also called van der Waals forces.
The statistical models confined by walls can be divided, in general, 
into two sets: the semi-infinite systems, in which at least 
one of the spatial coordinates is unconstrained by the walls,
and the fully-finite systems.
The applied methods and observed phenomena usually depend on this 
classification. 
Here, we shall restrict ourselves to the semi-infinite 
confining geometries, namely the polarizable interface (figure 1) 
and the two-slab geometry (figure 2).
The fully-finite geometries will be treated in a subsequent paper.

\begin{figure}[t]
\begin{center}
\includegraphics[scale=0.9]{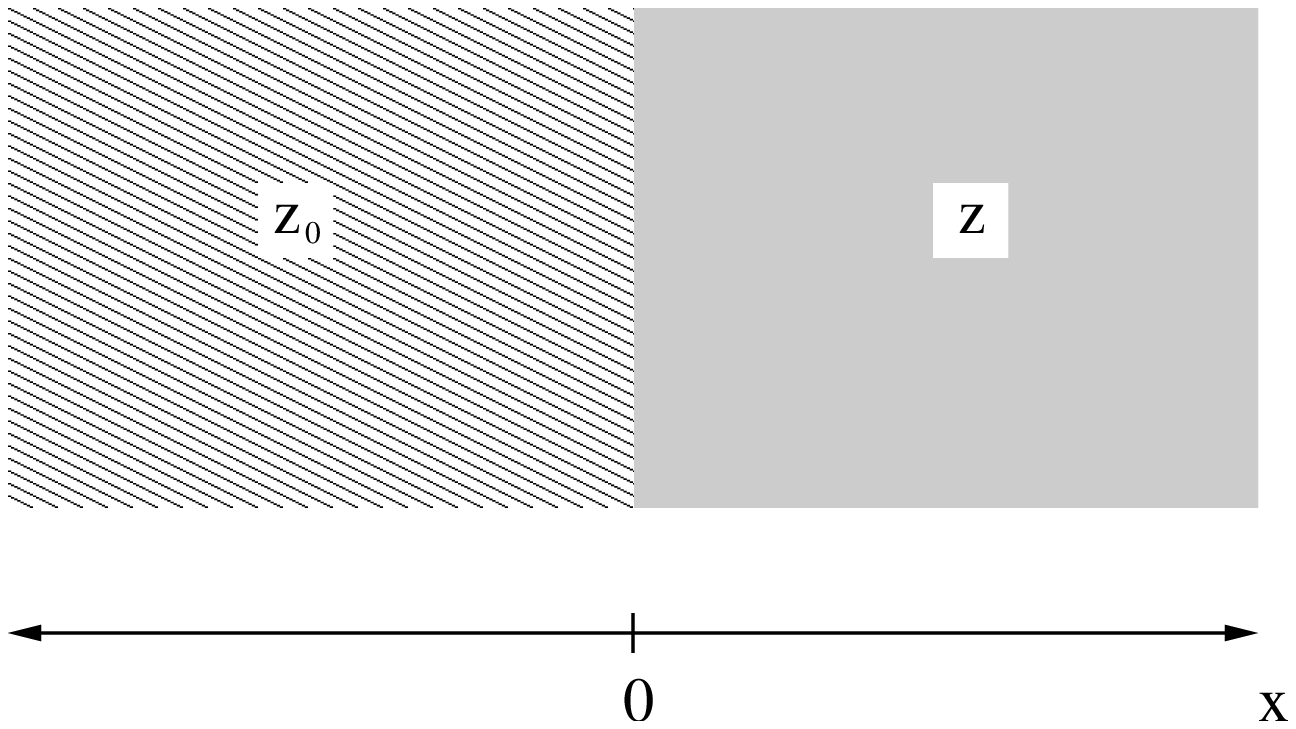}
\mbox{Figure 1. Polarizable interface.}
\end{center}
\end{figure}

\begin{figure}[b]
\begin{center}
\includegraphics[scale=0.9]{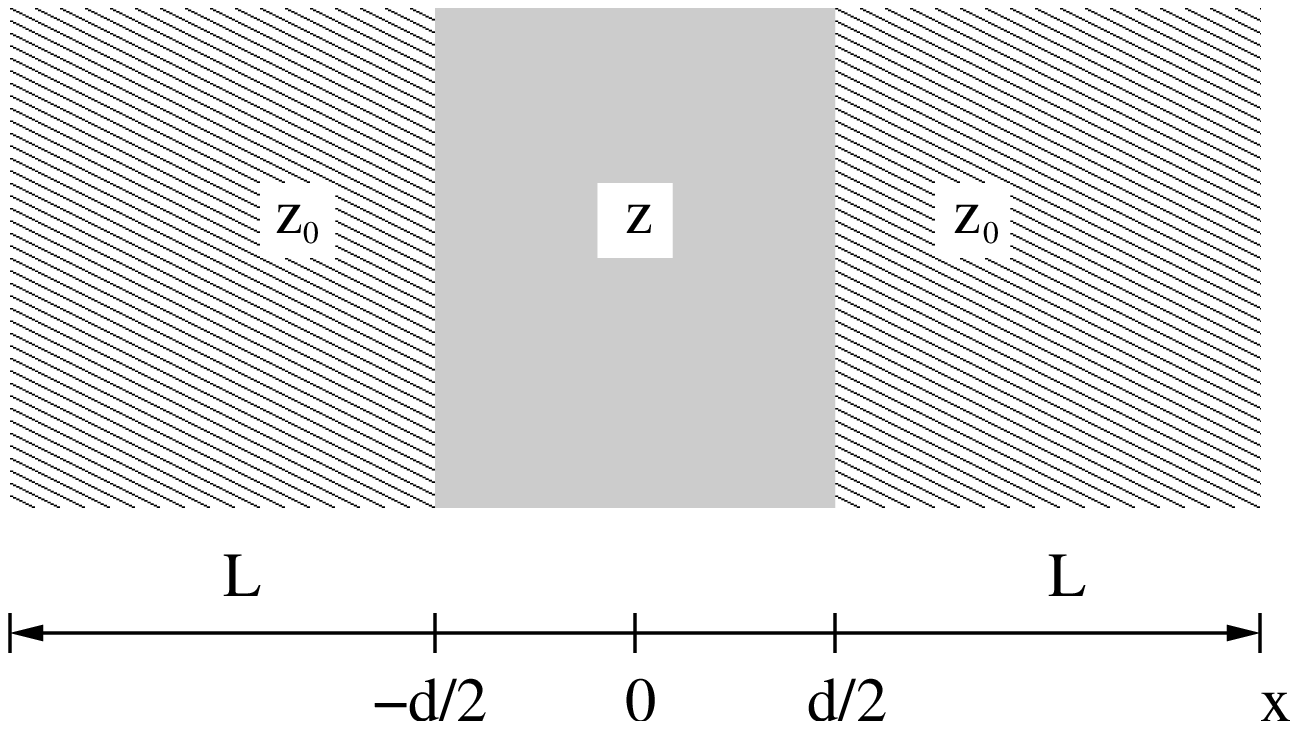}
\mbox{Figure 2. Two-slab geometry.}
\end{center}
\end{figure}

The studied models include the possibility of the presence 
of an electrolyte between the conducting walls. 
It is generally believed that the long-ranged Casimir force between 
the fluctuating walls is screened by this electrolyte.
Such an intuitively expected behaviour was supported longtime ago by 
an approximate Poisson-Boltzmann analysis of an electrolyte between 
two dielectric slabs \cite{Attard}.
A reasonable, but not rigorously justified, assumption was made: 
the electrostatic-field fluctuations in the walls, responsible for 
the long-ranged attractive Casimir force, can be treated separately 
from the pure image effects of the ``inert'' walls on 
the charged constituents of the electrolyte.
Hereinafter, we refer to this assumption as the ``separation hypothesis''.
Considering then the pure effect of images on the electrolyte particles, 
a repulsive long-ranged force of the same amplitude as the Casimir one, 
but in the opposite direction, occurs in the electrolyte.
Following the basic separation hypothesis, the two long-ranged forces 
cancel exactly with one another; we shall refer to this cancellation
phenomenon as the screening effect.
By the use of the separation hypothesis, this screening effect was 
also observed for an electrolyte between two conducting walls 
\cite{Tellez}.
The presence of the repulsive long-ranged force in the electrolyte
(modelled by the Coulomb gas) has been proven at arbitrary temperature, 
by using perfect-screening sum rules, for ideal-conductor 
\cite{Jancovici1} and ideal-dielectric \cite{Jancovici2} inert slab walls.
Interestingly, in the special case when $\epsilon=1$ (plain hard walls
separating the electrolyte from vacuum), there is neither an attractive 
Casimir force between the slabs nor a repulsive force in the electrolyte 
induced by the particle images in the inert walls 
(in fact, there are no images).
The corresponding field theory was developed in a series of works 
by Dean and Horgan \cite{Dean}.

In references \cite{Forrester1,Jancovici3}, and in a recent work
\cite{Buenzli} about the microscopic origin of universality in 
Casimir forces, the conducting walls in empty space are modelled by 
two slabs filled with a classical plasma of charged particles 
in thermal equilibrium.
At large distances between the two slabs, the Casimir force
(in the electrostatic regime) is retrieved as a result
of perfect-screening sum rules. 
A natural continuation of the strategy is to replace the ideal 
conductor walls by a microscopic plasma system in more general
physical situations.

The aim of the present paper is to study Casimir and related effects 
in model systems with geometries presented in figures 1 and 2,
in which the confining walls are made of a symmetric 
two-component plasma (Coulomb gas) of point-like particles with
$+/-$ unit charges in thermal equilibrium.
The Coulomb gas is chosen for the sake of its simplicity;
most of the obtained results can be easily generalized to
multi-component plasmas.
In the presence of an electrolyte (which is simply another Coulomb gas
with a different particle fugacity) between the walls, 
the microscopic description of the walls enables us to mimic coherently 
the effect of electrostatic fluctuations inside the walls and 
the image forces acting on the electrolyte particles, 
without any ad-hoc separation ansatz.
The wall-plasma is not required to be ideal, there might be a finite 
nonzero correlation length which describes the correlation effects among
the charged particles.

If there is no electrolyte in the region between the two slabs (figure 2), 
the Casimir force is obtained in the limit of a large wall 
separation, together with a leading correction term 
due to the non-ideality of the conductor wall.
If there is an electrolyte between the two slabs, the Casimir force is 
shown to be screened by this electrolyte, which confirms the validity of 
the separation (fluctuations-images) hypothesis.
Interestingly, the residual short-ranged force between the slabs
is always attractive.  
As a by-product of the treatment, we consider also the ideal-conductor 
regime of the walls when the correlation length of the charged particles 
forming the walls goes to zero.
The surface tension of the electrolyte turns out to be independent 
of the electrostatic field fluctuations in the ideal-conductor walls.
The analogous independence of the density profile and particle 
correlation functions was previously noted \cite{Cornu1,Forrester2}.

The spatial dimension of the studied models will be $\nu= 2, 3$.
In dimension $\nu$, the Coulomb potential $v$ at a spatial position
${\bf r}$, induced by a unit charge at the origin ${\bf 0}$, is
the solution of the Poisson equation
\begin{equation} \label{1.1}
\Delta v({\bf r}) = - s_{\nu} \delta({\bf r}) ,
\end{equation}
where $s_{\nu} = 2 \pi^{\nu/2}/\Gamma(\nu/2)$ is the surface area
of the $\nu$-dimensional unit sphere.
Explicitly,
\begin{equation} \label{1.2}
v({\bf r}) = \cases{-{\rm ln}(r/a)&  if $\nu=2$, \\
r^{2-\nu}/(\nu-2)& otherwise.}
\end{equation}
Here, $r=\vert {\bf r}\vert$ and $a$ is a free length scale.
The definition (\ref{1.1}) of the Coulomb potential implies
in the Fourier space the characteristic small-$k$ behaviour
${\hat v}({\bf k})\propto 1/k^2$ in any dimension.
This maintains many generic properties (like screening \cite{Martin})
of ``real'' 3D Coulomb systems.
The interaction energy of charged particles $\{ i, q_i=\pm 1\}$,
immersed in a homogeneous medium of dielectric constant = 1, is
$\sum_{i<j} q_i q_j v(\vert {\bf r}_i-{\bf r}_j \vert)$.

The models are treated in the grand canonical ensemble characterized
by the inverse temperature $\beta$ and by the couple
of equivalent (there is no external electrostatic potential)
particle fugacities $z_+({\bf r}) = z_-({\bf r}) = z({\bf r})$.
The whole domain $\Lambda$ on which the system is defined can be
separated into disjunct subdomains, 
$\Lambda = \cup_{\alpha} \Lambda^{(\alpha)}$.
Within the grand canonical formalism, each subdomain is
characterized by a constant fugacity, $z_q({\bf r}) = z^{(\alpha)}$ 
for ${\bf r}\in \Lambda^{(\alpha)}$, which may vary 
from one region to the other.
The choice $z^{(\alpha)}=0$ corresponds to a vacuum subdomain with
no particles allowed to occupy the space.
The grand partition function is defined by
\begin{equation} \label{1.3}
\fl \Xi = \sum_{N_+,N_-=0}^{\infty} \frac{1}{N_+! N_-!}
\int \prod_{i=1}^N \left[ \rmd^{\nu}r_i\, z_{q_i}({\bf r}_i) \right]
\exp\left[ -\beta \sum_{i<j} q_i q_j 
v(\vert {\bf r}_i-{\bf r}_j \vert) \right] ,
\end{equation}
where $N_+$ ($N_-$) is the number of positively (negatively) charged
particles and $N=N_++N_-$.
The averaged particle densities are generated from $\Xi$ in a standard 
way as functional derivatives with respect to the fugacity field.
At the one-particle level, one introduces the number density
of particles of one sign,
\begin{equation} \label{1.4}
n_q({\bf r}) = \left\langle \sum_i \delta_{q,q_i} 
\delta({\bf r}-{\bf r}_i) \right\rangle
= \frac{1}{\Xi} \frac{\delta \Xi}{\delta z_q({\bf r})} .
\end{equation}
Due to the charge symmetry, $n_+({\bf r}) = n_-({\bf r}) = n({\bf r})/2$
where $n$ is the total density of particles.
At the two-particle level, one introduces the two-body densities
\begin{eqnarray} 
n_{qq'}({\bf r},{\bf r}') & = & \left\langle \sum_{i\ne j} \delta_{q,q_i} 
\delta({\bf r}-{\bf r}_i) \delta_{q',q_j}  \delta({\bf r}'-{\bf r}_j)  
\right\rangle \nonumber \\
& = & \frac{1}{\Xi} \frac{\delta^2 \Xi}{\delta z_q({\bf r})
\delta z_{q'}({\bf r}')} . \label{1.5}
\end{eqnarray}
It is useful to consider also the (truncated) pair correlation functions
\begin{equation} \label{1.6}
h_{qq'}({\bf r},{\bf r}') = 
\frac{n_{qq'}({\bf r},{\bf r}')}{n_q({\bf r}) n_{q'}({\bf r}')} - 1 .
\end{equation}
For the case of point-like particles, the singularity of the Coulomb
potential (\ref{1.2}) at the origin often prevents the thermodynamic
stability against the collapse of positive-negative pairs of charges:
in 2D for small enough temperatures, in 3D for any temperature.
In particular, in 2D, the Boltzmann factor of a pair of $+/-$ unit charges, 
$r^{-\beta}$, is integrable at short distances provided that $\beta<2$. 
To cross the collapse point $\beta=2$, the pure Coulomb potential has to 
be regularized by a short-distance repulsion, e.g., a hard-core potential.
In 3D, the short-distance regularization is needed at any temperature
(except in the high-temperature limit described by the Debye-H\"uckel 
approximation which gives a thermodynamically stable description
for point-like charges).

Two methods are applied. 
We first use the Debye-H\"uckel method of the Coulomb-bond 
chain resummation within the density Mayer expansion, valid in the 
high-temperature limit and applicable to both 2D and 3D systems.
The standard approach dealing with inhomogeneous density profiles 
\cite{Jancovici4,Choquard} has to be supplemented by the first 
equation of the BGY hierarchy. 
We rather use the version developed in \cite{Kalinay,Jancovici5}
which does not require some additional information; we also introduce
some technicalities which substantially simplify the calculation of 
the Casimir forces.
We also use the Debye-H\"uckel method in the case of dielectric walls,
using the separation hypothesis which does not require any microscopic
model for these walls.
The second method applies exclusively to the 2D Coulomb gas
whose exact solution (Ursell functions) can be obtained just at 
the collapse temperature $\beta=2$ by a mapping onto the Thirring
model at the free-fermion point \cite{Cornu1,Cornu2}.
The free-fermion model is solvable also in many inhomogeneous situations
and the corresponding method can be adapted to the present problems.
The fact that, in 2D, one can verify whether some features of 
the mean-field behaviour predicted by the Debye-H\"uckel analysis 
persist also at a specific finite temperature is of importance.
We add that, based on a recent progress in the integrable $1+1$ 
dimensional Quantum Field Theory \cite{Zamolodchikov,Bajnok}, 
the thermodynamics of the 2D Coulomb gas was solved exactly 
in the whole stability region $\beta<2$, in the bulk \cite{Samaj2} as
well as for interface geometries with an ideal-conductor \cite{Samaj3} 
and an ideal-dielectric \cite{Samaj4} walls; however, for $\beta<2$,
the exact $n$-body densities are not known, and we shall not consider
the case $\beta<2$ in the present paper.
 
The paper is organized as follows.
In section 2, for the two-slab geometry (figure 2) with general dielectric 
walls, by using a phenomenological approach, it is shown, under
certain conditions, that the electric field fluctuations can be 
decoupled from the image-particle forces. 
The separation hypothesis is then tested on the microscopic models, 
first in  the Debye-H\"uckel limit (section 3) and then at 
the free-fermion point of the Thirring representation of 
the 2D Coulomb gas (section 4).
A brief recapitulation and concluding remarks are given in section 5.

\section{Phenomenology}
\subsection{The separation hypothesis}
A general argument in favour of the separation hypothesis has been given 
in the case of conducting walls \cite{Jancovici3}. 
After a minor modification, the same argument can be used 
for dielectric walls of arbitrary dielectric constant $\epsilon$. 
For the sake of completeness, we repeat the argument.

We consider, in $\nu$ dimensions, a Coulomb gas partially or
totally bounded by walls $W$ made of a dielectric with a dielectric
constant $\epsilon$. 
We want to take into account the thermal fluctuations of the walls. 
It happens that a detailed knowledge of the internal structure of 
the walls is not necessary; we only assume that this internal structure 
is described by a set of coordinates $\{{\bf R}\}$. 
Classical (non-quantum) statistical mechanics is supposed to be applicable. 

We shall need a correlation function defined for the wall system
supposed to be alone in space. 
Let $\Phi_W({\bf r})$ be the electric potential created by the walls 
at some point ${\bf r}$ and let 
$\langle \Phi_W({\bf r})\Phi_W({\bf r}')\rangle_0^{\rm T}$ 
be a truncated statistical average computed with the Boltzmann weight 
of the walls alone. 
This correlation function, when both ${\bf r}$ and ${\bf r}'$ are
outside the walls, can be computed by linear response theory, by a
simple adaptation of reference \cite{Janco}. 
Let us put a point charge $q$ at ${\bf r}'$. 
Its interaction with the walls results into the addition to
the Hamiltonian of a term $\delta H = q \Phi_W({\bf r}')$. 
The average electric potential at ${\bf r}$ is changed by 
$q G({\bf r},{\bf r}')$ with $G$ determined by macroscopic electrostatics. 
For instance, in the case of one plane wall, 
$G({\bf r},{\bf r}')=v({\bf r}-{\bf r}')+
[(1-\epsilon)/(1+\epsilon)]v({\bf r}-{\bf r}'^*)$ 
where $v$ is the Coulomb interaction (1.2) and 
${\bf r}'^*$ is the image of ${\bf r}'$.
 
Let $-G^*$ be that part of $G$ due to the walls (not to $q$ itself):
\begin{equation} \label{2.1}
-G^*({\bf r},{\bf r}') = G({\bf r},{\bf r}') - v({\bf r}-{\bf r}') .
\end{equation}
If $q$ is infinitesimal, linear response theory states that 
$-qG^*=-\beta\langle \Phi_W({\bf r})\delta H\rangle_0^{\rm T}$, 
therefore
\begin{equation} \label{2.2}
\beta \langle \Phi_W({\bf r}) \Phi_W({\bf r}') \rangle_0^{\rm T} =
G^*({\bf r},{\bf r}') .
\end{equation}
Furthermore, we can assume that the fluctuations of $\Phi_W({\bf r})$ 
are Gaussian, because, in a macroscopic description with a constant
$\epsilon$, the response of the walls to the external charge $q$ is
linear even for a non-infinitesimal charge (if the response of a spring
to an applied force is linear, it means that the spring is harmonic, and
its spontaneous thermal fluctuations will be Gaussian).  

We now come back to the full system Coulomb gas plus walls, the
partition function of which will be considered. 
Let $H_0(\{{\bf R}\})$ be the potential energy of the walls, 
$\{{\bf r}\}$ the set of the particle coordinates of of the Coulomb gas. 
The total potential energy of the system (Coulomb gas plus $W$) is 
\begin{equation} \label{2.3} 
H(\{{\bf r}\},\{{\bf R}\}) = \sum_i q_i \Phi_W({\bf r}_i,\{{\bf R}\}) +
\sum_{i<j} q_i q_j v({\bf r}_i-{\bf r}_j) + H_0(\{{\bf R}\}) ,
\end{equation}
where $q_i$ and ${\bf r}_i$ are the charge and coordinates of the $i$th
particle of the Coulomb gas. 
Let $\rmd\Gamma$ the phase space element (after integration 
on the momenta) of the Coulomb gas. 
The partition function $Z$ of the total system can be written as 
\begin{equation} \label{2.4}
\fl Z = Z_0 \int \rmd \Gamma \left\langle \exp\left[ 
- \beta \sum_i q_i\Phi_W({\bf r}_i,\{{\bf R}\}) \right] \right\rangle_0 
\exp\left[ - \beta \sum_{i<j} q_i q_j v({\bf r}_i-{\bf r}_j) \right] ,
\end{equation} 
where $\langle \cdots \rangle_0$ means the average over $\{{\bf R}\}$ 
with the weight $\exp[-\beta H_0]$, and $Z_0$ is the partition function 
of the walls alone. 
Since the fluctuations of $\Phi_W$ are Gaussian and
$\langle \Phi_W({\bf r}) \rangle_0 = 0$,
\begin{eqnarray}
\fl \left\langle \exp\left[ - \beta \sum_iq_i \Phi_W({\bf r}_i,\{{\bf R}\})
\right] \right\rangle_0 = \exp\left[ \case12 \beta^2\sum_{i,j} q_i q_j 
\langle \Phi_W({\bf r}_i)\Phi_W({\bf r}_j)\rangle_0^{\rm T} \right] 
\nonumber \\
\lo= \exp\left[ \case12 \beta \sum_{i,j} q_i q_j G^*({\bf r}_i,{\bf r}_j)
\right] . \label{2.5}
\end{eqnarray}
Thus, the partition function can be factorized (separated) as
\begin{equation} \label{2.6}
Z=Z_{\rm eff} Z_0 ,
\end{equation}
where $Z_{\rm eff}$ is the partition function of the Coulomb gas 
computed with the effective Hamiltonian
\begin{equation} \label{2.7}
H_{\rm eff} = - \frac{1}{2} \sum_i q_i^2 G^*({\bf r}_i,{\bf r}_i) +
\sum_{i<j} q_i q_j G({\bf r}_i,{\bf r}_j) .
\end{equation}
$H_{\rm eff}$ is the standard Hamiltonian of a Coulomb gas in presence
of inert dielectric walls. 
For instance, in the case of one plane dielectric wall, 
the first term of $H_{\rm eff}$ is the interaction of
each particle with its own image, while the second term is the
interaction of each particle with the other ones and the images of the
other ones. 

Let $A(\{{\bf r}\})$ be any microscopic quantity 
which does not depend on $\{{\bf R}\}$. 
Using the same kind of steps as for the calculation of
the partition function $Z$, one can show that this average can be
calculated with $H_{\rm eff}$ only:
\begin{equation} \label{2.8}
\langle A \rangle=\frac{1}{Z_{\rm eff}}\int \rmd \Gamma 
\exp(-\beta H_{\rm eff}) A .
\end{equation}
In particular, the densities and many-body correlations in the Coulomb
gas can be calculated with $H_{\rm eff}$ and therefore are unaffected by
the fluctuations in the walls. 

On the contrary, the total partition function (\ref{2.6}) does involve
$Z_0$, and therefore the fluctuations in the walls enter the total free
energy; if a wall is displaced, the variation of the free energy, and
therefore the force acting on this wall, will be affected by the
fluctuations in the walls. 
 
In the above, simple manipulations allow to replace the partition
function $Z_{\rm eff}$ of the Coulomb gas by the corresponding grand 
partition function $\Xi_{\rm eff}$ also computed with 
the Hamiltonian $H_{\rm eff}$.
Therefore, the grand potential of the total system is
\begin{equation} \label{2.9}
\Omega=\Omega_{\rm eff} + F_0 ,
\end{equation}
where $\Omega_{\rm eff}$ is the grand potential of the Coulomb gas 
computed with the Hamiltonian $H_{\rm eff}$ and $F_0$ is the free energy 
of the wall system alone in space.
  
Furthermore, we have considered the case when the Coulomb gas is made of
charged particles in vacuum. 
A more realistic model for an electrolyte
takes into account the solvent by picturing it as a continuous
dielectric medium with a dielectric constant $\epsilon_S$. 
Then, in the above, $\epsilon$ means the ratio $\epsilon_W/\epsilon_S$ 
where now $\epsilon_W$ is the dielectric constant of the walls.

In the limit $\epsilon\rightarrow\infty$, the results about conducting
walls \cite{Jancovici3} are retrieved.

\subsection{Casimir force between two dielectric walls}
In the present subsection, we give a simple argument for computing 
the Casimir force between two parallel dielectric walls separated by vacuum. 
We work in $\nu$ dimensions. 
The walls, perpendicular to the $x$ axis, modelled as media with a dielectric
constant $\epsilon$, occupy the regions $x>d/2$ and $x<-d/2$.
The slab $-d/2 < x < d/2$ is empty. 
Each point ${\bf r}$ is defined by its Cartesian coordinates 
$(x,{\bf r}^{\perp})$ where ${\bf r}^{\perp}$ is the $(\nu -1)$ 
dimensional vector perpendicular to the $x$ axis.
We shall compute the average $xx$ component of the Maxwell stress
tensor \cite{Jackson} in the empty region.

Let $G({\bf r},{\bf r}')$ be the electric potential at ${\bf r}$ when a
unit charge is placed  at the source point ${\bf r}'$ in the empty region. 
Macroscopic electrostatics gives for $G$ the equations  
\begin{equation} \label{2.10}
\Delta G({\bf r},{\bf r}') =\cases{
-s_{\nu} \delta({\bf r}-{\bf r}') & if $- d/2 < x < d/2 $ , \\
0& if $x < - d/2$ or $x > d/2$ ,} 
\end{equation}
with the conditions that $G$ and $\epsilon(\partial G/\partial x)$ 
[with $\epsilon=1$ if $- d/2 < x < d/2$] be continuous at $x = - d/2$ and
$x = d/2$, and that $G\rightarrow 0$ when $x\rightarrow\pm\infty$. 
The standard technique is to use the Fourier transform $\hat{G}$ in the 
${\bf r}^{\perp}$ space, defined as in (\ref{3.15}). 
A derivation similar to the ones in section 3 gives for $G^*$, 
defined by (\ref{2.1}), the Fourier transform $\hat{G}^*$  
\begin{eqnarray} 
\fl \hat{G}^* = - \frac{s_{\nu}}{l} \left\{ \cosh[l(x-x')] 
\left[\left(\frac{1+\epsilon}{1-\epsilon}\right)^2 {\rm e}^{2ld}-1 \right]^{-1}
\right. \nonumber \\
\left. + \cosh[l(x+x')]  \left[
\left( \frac{1+\epsilon}{1-\epsilon} \right) {\rm e}^{ld}-
\left( \frac{1-\epsilon}{1+\epsilon} \right) {\rm e}^{-ld} \right]^{-1} 
\right\} . \label{2.11} 
\end{eqnarray}

The $xx$ component $T_{xx}$ of the average Maxwell stress tensor at
${\bf r}$ is, from (\ref{2.2}) with 
$\langle \Phi_W({\bf r}) \rangle = 0$,
\begin{equation} \label{2.12}
T_{xx}({\bf r})=\frac{\beta^{-1}}{2s_{\nu}}\left[\frac{\partial^2}
{\partial x \partial x'}-
\nabla_{{\bf r}^{\perp}}\cdot\nabla_{{\bf r}'^{\perp}}\right]
G^*({\bf r},{\bf r}')\Big\vert_{{\bf r}'={\bf r}} .
\end{equation}
When $G^*$ is expressed in terms of its Fourier transform (\ref{2.11}),
(\ref{2.12}) becomes
\begin{eqnarray} 
\fl \beta T_{xx} = \frac{s_{\nu -1}}{(2\pi)^{\nu -1}}\int_0^{\infty}
\rmd l \, l^{\nu -1} \left[ \left( \frac{1+\epsilon}{1-\epsilon} \right)^2
\rme^{2ld}-1 \right]^{-1} \nonumber \\
\lo= \frac{s_{\nu -1}}{(2\pi)^{\nu -1}(2d)^{\nu}}
\left(\frac{1-\epsilon}{1+\epsilon}\right)^2\Gamma(\nu)
\Phi\left(\left(\frac{1-\epsilon}{1+\epsilon}\right)^2,\nu,1\right) ,
\label{2.13}
\end{eqnarray}
where $\Phi$ is the Lerch function \cite{Gradshteyn}
\begin{equation} \label{2.14}
\Phi(z,\nu,1)=\sum_{n=1}^{\infty}\frac{z^{n-1}}{n^{\nu}} .
\end{equation}

$-T_{xx}$ is independent of ${\bf r}$. 
It is the force per unit area, measured along the $x$-axis, 
which acts on the right wall. 
This force is attractive. 
It is related to the free energy of the wall system $F_0$ introduced 
in section 2.1 by
\begin{equation} \label{2.15} 
-T_{xx} = - \frac{1}{{\mathcal A}} \frac{\partial F_0}{\partial d} ,
\end{equation}
where ${\mathcal A}$ is the area of each wall.

If $\nu=3$, (2.13) is in agreement with Lifshitz \cite{Lifshitz}. 
In the limit of conducting walls $\epsilon\rightarrow\infty$, it reduces to 
$\beta T_{xx}=\zeta(3)/(8\pi d^3)$, where $\zeta$ is the Riemann
zeta function.
This is indeed the correct electrostatic part of the Casimir force 
between conducting walls in the classical limit \cite{Schwinger,Forrester1}.

If $\nu=2$, in the limit of conducting walls, 
$\beta T_{xx} = \pi/(24 d^2)$, 
in agreement with a previous result \cite{Forrester1}. 
 
\section{Debye-H\"uckel theory}
The weak-coupling (high-temperature) limit of a Coulomb system 
is described by the Debye-H\"uckel theory.

\subsection{General formalism}
We first consider the bulk regime when the Coulomb system is 
defined in the whole space, $\Lambda=R^{\nu}$, and the fugacities 
$\{ z_q\}$ of particles with charges $q\in Q$ as well as 
the corresponding particle densities $\{ n_q \}$ are constant in space. 
A consistent way of deriving the Debye-H\"uckel theory is to start
with the diagrammatic Mayer expansion in density.
Summing up the chain diagrams leads to the renormalized bonds,
and the cluster integrals converge in the renormalized format
(see, for instance, \cite{Jancovici5,Samaj2}).
This is equivalent to writing the Ornstein-Zernicke (OZ) equations
for the pair correlation function (\ref{1.6}) with the direct
correlation function replaced by $-\beta$ times the corresponding
interaction potential:
\begin{eqnarray}
\fl h_{q_1 q_2}({\bf r}_1,{\bf r}_2) = - \beta q_1 q_2
v(\vert {\bf r}_1-{\bf r}_2 \vert) \nonumber \\
+ \sum_{q_3} \int_{\Lambda} \rmd^{\nu}r_3
\left[ -\beta q_1 q_3 v(\vert {\bf r}_1-{\bf r}_3 \vert) \right]
n_{q_3}({\bf r}_3) h_{q_3 q_2}({\bf r}_3,{\bf r}_2) \label{3.1}
\end{eqnarray}
The solution is of the form
\begin{equation} \label{3.2}
h_{q_1 q_2}({\bf r}_1,{\bf r}_2) = - \beta q_1 q_2
G({\bf r}_1,{\bf r}_2) ,
\end{equation}
where $G$ obeys the integral equation
\begin{equation} \label{3.3}
G({\bf r}_1,{\bf r}_2) = v(\vert {\bf r}_1-{\bf r}_2 \vert)
- \frac{\kappa^2}{s_{\nu}} \int_{\Lambda} \rmd^{\nu} r_3
v(\vert {\bf r}_1-{\bf r}_3 \vert) G({\bf r}_3,{\bf r}_2) .
\end{equation}
Here, $\kappa$ is the inverse Debye length defined by
$\kappa^2 = s_{\nu} \beta \sum_q n_q q^2$.
The integral equation (\ref{3.3}) can be transformed into
a differential equation by taking the Laplacian 
with respect to ${\bf r}_1$.
In this way one arrives at the usual Debye-H\"uckel equation
\begin{equation} \label{3.4}
\left[ \Delta_1 - \kappa^2 \right] G({\bf r}_1,{\bf r}_2)
= - s_{\nu} \delta({\bf r}_1 - {\bf r}_2)
\end{equation} 
which has to be supplemented by the vanishing boundary conditions
for the screened Coulomb potential $G$ when 
$\vert {\bf r}_1-{\bf r}_2 \vert \to \infty$.
The bulk density-fugacity relationship can be obtained by using
the formula
\begin{equation} \label{3.5}
n_q = z_q \exp\left\{ \frac{\beta q^2}{2} \lim_{r\to 0}
\left[ -G(r) + v(r) \right] \right\} .
\end{equation}
The specific grand partition function can be calculated
from the standard relation
\begin{equation} \label{3.6}
n_q = z_q \frac{\partial}{\partial z_q}
\frac{{\rm ln}\Xi}{\vert \Lambda \vert}
\end{equation}
with the condition ${\rm ln}\Xi = 0$ at $\{ z_q=0 \}$.
The bulk pressure $p$ is given by $\beta p = \ln\Xi/\vert\Lambda\vert$.

In the 2D Coulomb gas, the screened Coulomb potential 
is found in the form
\begin{equation} \label{3.7}
G(r) = K_0(\kappa r) , \quad \quad \kappa =(2\pi\beta n)^{1/2}
\quad \quad (\nu=2) ,
\end{equation}
where $K_0$ is a modified Bessel function.
Using the small-$x$ expansion of $K_0(x)$ \cite{Gradshteyn} 
\begin{equation} \label{3.8}
K_0(x) = - {\rm ln}\left( \frac{x}{2} \right) - C
+ O(x^2{\rm ln}x)
\end{equation}
with $C$ being the Euler's constant, the density-fugacity
relationship (\ref{3.5}) reads
\begin{equation} \label{3.9}
\frac{n^{1-\beta/4}}{z a^{\beta/2}} =
2 \left( \frac{\pi \beta}{2} \right)^{\beta/4}
\exp\left( \frac{\beta C}{2} \right) .
\end{equation}
The bulk pressure is given by $\beta p = (1-\beta/4) n$.
This equation of state turns out to be valid for any $\beta\le 2$.

In the 3D Coulomb gas, the screened Coulomb potential 
is found in the form
\begin{equation} \label{3.10}
G(r) = \frac{\exp(-\kappa r)}{r} , \quad \quad \kappa =(4\pi\beta n)^{1/2}
\quad \quad (\nu=3)
\end{equation}
The density-fugacity relationship then reads
\begin{equation} \label{3.11}
n = 2 \left[ z + \sqrt{2\pi} (\beta z)^{3/2} \right] .
\end{equation}
The bulk pressure is given by $\beta p = n - \kappa^3/(24 \pi)$.

The above Debye-H\"uckel theory can be straightforwardly extended
to inhomogeneous situations when the system domain $\Lambda$ 
is separated into disjunct subdomains, 
$\Lambda=\cup_{\alpha} \Lambda^{(\alpha)}$.
Each subdomain is characterized by constant fugacities,
$z_q({\bf r}) = z_q^{(\alpha)}$ for ${\bf r}\in \Lambda^{(\alpha)}$,
and the corresponding bulk densities $n_q^{(\alpha)}$.
At the lowest level, the densities in the OZ equations (\ref{3.1}) 
are taken constant (equal to the bulk ones) within each of the subdomains, 
$n_q({\bf r}) = n_q^{(\alpha)}$ for ${\bf r}\in \Lambda^{(\alpha)}$.
The solution of the OZ equations is again of the form (\ref{3.2}), 
where the screened Coulomb potential $G$ obeys the integral equation
\begin{equation} \label{3.12}
G({\bf r}_1,{\bf r}_2) = v(\vert {\bf r}_1-{\bf r}_2 \vert)
- \sum_{\alpha} \frac{\kappa_{\alpha}^2}{s_{\nu}} 
\int_{\Lambda^{(\alpha)}} \rmd^{\nu} r_3 
v(\vert {\bf r}_1-{\bf r}_3 \vert) G({\bf r}_3,{\bf r}_2) .
\end{equation}
Here, $\kappa_{\alpha}$ is the inverse Debye length for the
subdomain $\Lambda^{(\alpha)}$ defined by
$\kappa_{\alpha}^2 = s_{\nu} \beta \sum_q n_q^{(\alpha)} q^2$.
Equivalently,
\begin{equation} \label{3.13}
\left[ \Delta_1 - \kappa_{\alpha}^2 \right] G({\bf r}_1,{\bf r}_2)
= - s_{\nu} \delta({\bf r}_1-{\bf r}_2), \quad \quad
{\bf r}_1 \in \Lambda^{(\alpha)} ,
\end{equation}
where the spatial position of the source point ${\bf r}_2$ is arbitrary.
Since, according to (\ref{3.12}), $G$ is proportional to $v$,
these equations must be supplemented by the usual electrostatic
conditions at each subdomain boundary $\partial \Lambda^{(\alpha)}$:
$G$ and its normal derivative with respect to the boundary surface
$\partial_{\perp}G$ are continuous at $\partial \Lambda^{(\alpha)}$.
The leading $\beta$-correction to the particle densities in a given
subdomain is determined by the inhomogeneous counterpart of 
equation (\ref{3.5}),
\begin{equation} \label{3.14}
n_q^{(\alpha)}({\bf r}) = z_q^{(\alpha)} \exp \left\{
\frac{\beta q^2}{2} \lim_{{\bf r}'\to {\bf r}} 
\left[ - G({\bf r},{\bf r}') + v({\bf r},{\bf r}') \right] \right\}
\end{equation}
where ${\bf r}\in \Lambda^{(\alpha)}$.

\subsection{Polarizable interface}
The model is an infinite Coulomb gas of point unit charges $q=\pm 1$,
separated by an interface (perpendicular to the $x$ axis and localized
at $x=0$) into two half spaces $x>0$ and $x<0$ (see figure 1). 
Each point ${\bf r}\in R^{\nu}$ is thus defined by Cartesian coordinates
${\bf r} = (x,{\bf r}^{\perp})$, where ${\bf r}^{\perp}$ is the
$(\nu-1)$ dimensional vector component perpendicular to the $x$ axis. 
The particle fugacities are, in general, different on each side of the 
interface, $z_q({\bf r})=z$ when $x>0$ and $z_q({\bf r})=z_0$ when $x<0$.
The corresponding bulk (total number) densities and inverse Debye lengths
will be denoted by $(n,\kappa)$ and $(n_0,\kappa_0)$.
Since the system is translationally invariant in the perpendicular
${\bf r}^{\perp}$-subspace, the standard technique is to use the
Fourier transform
\begin{equation} \label{3.15}
G({\bf r}_1,{\bf r}_2) = \int \frac{\rmd^{\nu-1}l}{(2\pi)^{\nu-1}}
{\hat G}(x_1,x_2;{\bf l}) \exp\left\{ \rmi {\bf l}\cdot
({\bf r}_1^{\perp}-{\bf r}_2^{\perp}) \right\}
\end{equation}

We want to solve equations (\ref{3.13}) when, for instance, the source
point ${\bf r}_2 = (x_2,{\bf r}^{\perp}_2)$ has the coordinate $x_2>0$.
In terms of the Fourier component ${\hat G}$ one obtains ordinary
differential equations in one variable:
\numparts
\begin{eqnarray}
\left( \frac{\partial^2}{\partial x_1^2} - \kappa^2 - l^2 \right)
{\hat G}(x_1,x_2;{\bf l}) = - s_{\nu} \delta(x_1-x_2) , \qquad
& (x_1>0) , \label{3.16a} \\
\left( \frac{\partial^2}{\partial x_1^2} - \kappa_0^2 - l^2 \right)
{\hat G}(x_1,x_2;{\bf l})  =  0 , & (x_1<0) . \label{3.16b}
\end{eqnarray}
\endnumparts
We look for a solution, vanishing at $x_1\to\pm\infty$, of the form
\numparts
\begin{eqnarray}
\fl {\hat G} =  \frac{s_{\nu}}{2\sqrt{\kappa^2+l^2}}
\exp\left( - \sqrt{\kappa^2+l^2} \vert x_1-x_2 \vert \right)
& \nonumber \\
+ A_l \exp\left[ - \sqrt{\kappa^2+l^2} (x_1+x_2) \right] ,
\qquad & ( x_1>0 ), \label{3.17a} \\
\fl {\hat G} =  B_l \exp \left( \sqrt{\kappa_0^2+l^2} x_1
- \sqrt{\kappa^2 + l^2} x_2 \right) , & ( x_1<0 ) . \label{3.17b}
\end{eqnarray}
\endnumparts
The constants $A_l$ and $B_l$ are determined by the conditions that
${\hat G}$ and $\partial {\hat G}/\partial x_1$ are continuous at
$x_1=0$.
The result is
\numparts
\begin{eqnarray}
A_l = \frac{s_{\nu}}{2\sqrt{\kappa^2+l^2}}
\frac{\sqrt{\kappa^2+l^2}-\sqrt{\kappa_0^2+l^2}}{\sqrt{\kappa^2+l^2}
+\sqrt{\kappa_0^2+l^2}} , \label{3.18a} \\
B_l = s_{\nu} \frac{1}{\sqrt{\kappa^2+l^2}+\sqrt{\kappa_0^2+l^2}} .
\label{3.18b} 
\end{eqnarray}
\endnumparts

The leading $\beta$-correction to the constant particle density can
now be calculated for $x>0$ by using a linearization in (\ref{3.14}).
Considering $x_1, x_2 > 0$, the first term on the rhs of (\ref{3.17a})
contributes to the bulk density-fugacity relationship, while the
second term implies
\begin{equation} \label{3.19}
n(x) = n \left[ 1 - \frac{\beta}{2} \int 
\frac{\rmd^{\nu-1}l}{(2\pi)^{\nu-1}} A_l
\exp\left( - 2 \sqrt{\kappa^2+l^2} x \right) \right] .
\end{equation}
More explicitly, one has
\begin{equation} \label{3.20}
\fl n(x)-n = \frac{\kappa^2 s_{\nu-1}}{4 (2\pi)^{\nu-1}}
\int_0^{\infty} \frac{\rmd l \, l^{\nu-2}}{\sqrt{\kappa^2+l^2}}
\frac{\sqrt{\kappa_0^2+l^2}-\sqrt{\kappa^2+l^2}}{\sqrt{\kappa_0^2+l^2}
+\sqrt{\kappa^2+l^2}} \exp\left( - 2 \sqrt{\kappa^2+l^2} x \right)
\end{equation}
for the density profile at $x>0$.
The same method applies when the source point 
${\bf r}_2 = (x_2,{\bf r}_2^{\perp})$ has the component $x_2<0$,
with the final result
\begin{equation} \label{3.21}
\fl n_0(x)-n_0 = \frac{\kappa_0^2 s_{\nu-1}}{4 (2\pi)^{\nu-1}}
\int_0^{\infty} \frac{\rmd l \, l^{\nu-2}}{\sqrt{\kappa_0^2+l^2}}
\frac{\sqrt{\kappa^2+l^2}-\sqrt{\kappa_0^2+l^2}}{\sqrt{\kappa^2+l^2}
+\sqrt{\kappa_0^2+l^2}} \exp\left( 2 \sqrt{\kappa_0^2+l^2} x \right)
\end{equation}
for the density profile at $x<0$.

For the present geometry, the grand potential 
$\Omega = - \beta^{-1} \ln \Xi$ is the sum of a volume part
and a surface part,
\begin{equation} \label{3.22}
\Omega = - \frac{\vert \Lambda \vert}{2} \left[ p(z) + p(z_0) \right]
+ \vert \partial\Lambda \vert \gamma(z,z_0),
\end{equation}
where $\vert \partial\Lambda \vert$ stands for the interface area
and $\gamma$ for the surface tension.
The total number of particles is given by
\begin{equation} \label{3.23}
N = z \frac{\partial (-\beta\Omega)}{\partial z}
+ z_0 \frac{\partial (-\beta\Omega)}{\partial z_0} .
\end{equation}
The surface parts of this relation read \cite{Cornu1}
\numparts
\begin{eqnarray}
z \frac{\partial (-\beta \gamma)}{\partial z}
= \int_0^{\infty} \rmd x \left[ n(x) - n \right] , \label{3.24a} \\
z_0 \frac{\partial (-\beta \gamma)}{\partial z_0}
= \int_{-\infty}^0 \rmd x \left[ n_0(x) - n_0 \right] . \label{3.24b}
\end{eqnarray}
\endnumparts
The surface tension is zero in the homogeneous bulk regime,
\begin{equation} \label{3.25}
\beta \gamma(z,z_0) = 0 \quad \mbox{when $z=z_0$.}
\end{equation}
At the lowest order in $\beta$, $z\partial_z$ and $z_0\partial_{z_0}$
can be replaced respectively by $n\partial_n$ and $n_0\partial_{n_0}$.
Having at one's disposal the density profiles, equations (\ref{3.20}) and
(\ref{3.21}), after some algebra one finally arrives at
\begin{equation} \label{3.26}
\beta \gamma = \frac{s_{\nu-1}}{4 (2\pi)^{\nu-1}} \int_0^{\infty}
\rmd l \, l^{\nu-2} \, {\rm ln} \left[
\frac{\left( \sqrt{\kappa_0^2+l^2}+\sqrt{\kappa^2+l^2} \right)^2}{
4 \sqrt{\kappa_0^2+l^2} \sqrt{\kappa^2+l^2}} \right] .
\end{equation}
Note that the argument of the logarithm behaves like
$1+ O(1/l^4)$ for large $l$, so the integral is finite in both
$\nu=2$ and $3$ dimensions of interest.
When there are no particles at the half-space $x<0$, i.e. $\kappa_0=0$,
the relations
\begin{equation} \label{3.27}
\beta \gamma^{(\rm vac)}(\kappa) = \cases{
\frac{\kappa}{8\pi}(4-\pi) , & $\nu=2$ \\
\frac{\kappa^2}{32 \pi} (2 \ln 2 -1) , & $\nu=3$}
\end{equation}
define the surface tension of the plasma in contact with a vacuum
(plain hard wall with $\epsilon=1$) \cite{Russier}.

We now aim at studying the formula (\ref{3.26}) in the ideal-conductor 
limit when the correlation length of the charged particles,
say in the half-space $x<0$, goes to zero, i.e. $\kappa_0\to\infty$.
The analysis depends on the spatial dimension.
\begin{itemize}

\item In 2D, the large-$\kappa_0$ expansion of (\ref{3.26})
results in a decomposition
\begin{equation} \label{3.28}
\beta \gamma = \frac{\kappa_0}{8\pi}(4-\pi) - \frac{\kappa}{8}
+\beta \gamma_{\rm mix}
\end{equation}
where
\begin{equation} \label{3.29}
\beta\gamma_{\rm mix} = \frac{\kappa_0}{2\pi} \int_0^{\infty}
\rmd k \, {\rm ln}\left[ \frac{\sqrt{1+k^2}+
\sqrt{(\kappa/\kappa_0)^2+k^2}}{\sqrt{1+k^2}+k} \right] .
\end{equation}
On the rhs of (\ref{3.28}), the first term corresponds to the
vacuum surface tension of the Coulomb gas in the half-space $x<0$
(owing to $\kappa_0>>\kappa$, the relatively sparse Coulomb gas
localized at $x>0$ is in fact seen as vacuum) and the second term 
corresponds to the surface tension of the Coulomb gas in the half-space 
$x>0$ in the presence of an inert ideal-conductor wall \cite{Samaj3}. 
The last term, defined by (\ref{3.29}), represents the ``mixing'' 
of the two surface tensions. 
It can be further simplified, via successive integrations by parts, 
to the form
\begin{equation} \label{3.30}
\beta \gamma_{\rm mix} = \frac{\kappa_0}{2\pi}
\left\{ \frac{\kappa}{\kappa_0} {\bf E}\left[ 1- 
\left( \frac{\kappa_0}{\kappa}\right)^2 \right] - 1 \right\} ,
\end{equation}
where ${\bf E}$ is a complete elliptic integral.
In the limit $\kappa/\kappa_0\to 0$, one has
\begin{equation} \label{3.31}
\frac{\beta\gamma_{\rm mix}}{\kappa} = 
- \frac{1}{4\pi} \frac{\kappa}{\kappa_0} {\rm ln}\left(
\frac{\kappa}{\kappa_0} \right) + O\left( \frac{\kappa}{\kappa_0} \right) ,
\end{equation}
so that the mixing surface tension vanishes in the ideal-conductor limit.
This means that the electrostatic field fluctuations in the wall do not 
modify the surface tension $-\kappa/8$ calculated with an inert 
ideal-conductor wall.

\item In 3D, the integral in equation (\ref{3.26}) 
can be evaluated explicitly,
\begin{equation} \label{3.32}
\beta\gamma = \frac{1}{16\pi} \left[ - \frac{1}{2} (\kappa_0-\kappa)^2
+ \kappa_0^2 {\rm ln}\left( \frac{2\kappa_0}{\kappa_0+\kappa} \right)
+ \kappa^2 {\rm ln}\left( \frac{2\kappa}{\kappa_0+\kappa} \right) \right] .
\end{equation}
In the limit $\kappa/\kappa_0\to 0$,
\begin{equation} \label{3.33}
\beta \gamma = \frac{\kappa_0^2}{32\pi} ( 2 {\rm ln} 2 - 1 )
+\frac{\kappa^2}{16\pi} {\rm ln}\left( \frac{2\kappa}{\kappa_0} \right)
+\beta \gamma_{\rm mix} ,
\end{equation}
where the mixing surface tension is given by the expansion formula
\begin{equation} \label{3.34}
\frac{\beta\gamma_{\rm mix}}{\kappa^2} = - \frac{1}{12\pi} 
\left( \frac{\kappa}{\kappa_0} \right)
+ O\left[ \left( \frac{\kappa}{\kappa_0} \right)^2 \right] . 
\end{equation}
Like in 2D, the first term on the rhs of (\ref{3.33}) is the
vacuum surface tension of the Coulomb gas in the half-space $x<0$.
The second term, the surface tension of the Coulomb gas in the
half-space $x>0$ due to the conductor wall, diverges in 
the ideal-conductor limit.
On the other hand, the mixing surface tension vanishes in the
ideal-conductor limit.
\end{itemize}

\subsection{Two-slab geometry}
The model consists of two slabs (see figure 2) modelled by Coulomb gases
with the same bulk characteristics $(n_0,\kappa_0)$.
The slabs have thickness $L$ and are separated by a distance $d$:
\begin{eqnarray}
\Lambda_- = \left\{ {\bf r}=(x,{\bf r}^{\perp}) \vert
x\in X_- = \langle -d/2-L,-d/2\rangle, {\bf r}^{\perp} \in R^{\nu-1}
\right\} , \nonumber \\
\Lambda_+ = \left\{ {\bf r}=(x,{\bf r}^{\perp}) \vert
x\in X_+ = \langle d/2,d/2+L\rangle, {\bf r}^{\perp} \in R^{\nu-1}
\right\} . \nonumber 
\end{eqnarray}
There is an electrolyte, modelled simply by another Coulomb gas
with different bulk characteristics $(n,\kappa)$, in the region
$(X_0 = \langle -d/2,d/2 \rangle) \times R^{\nu-1}$ between the slabs;
the choice $n=\kappa=0$ corresponds to vacuum between the slabs.
The remaining external space to the slabs is supposed to be vacuum.
The thickness $L$ of the slabs is large enough to screen the boundary
effects at $x=\pm d/2$, so that the opposite boundaries have the
density characteristics identical to those close to a plain hard wall.

Let the source point ${\bf r}_2 = (x_2,{\bf r}_2^{\perp})$
have the coordinate $x_2\in X_+$.
In terms of the Fourier component ${\hat G}$, 
equations (\ref{3.13}) take the form
\numparts
\begin{eqnarray}
\left( \frac{\partial^2}{\partial x_1^2} - \kappa_0^2 - l^2 \right)
{\hat G}(x_1,x_2;{\bf l}) = - s_{\nu} \delta(x_1-x_2),
\qquad & (x_1\in X_+), \label{3.35a} \\
\left( \frac{\partial^2}{\partial x_1^2} - \kappa^2 - l^2 \right)
{\hat G}(x_1,x_2;{\bf l}) = 0,
& (x_1\in X_0), \label{3.35b} \\
\left( \frac{\partial^2}{\partial x_1^2} - \kappa_0^2 - l^2 \right)
{\hat G}(x_1,x_2;{\bf l}) = 0,
& (x_1\in X_-). \label{3.35c}
\end{eqnarray}
\endnumparts
In the limit $L\to\infty$, we look for a solution, 
vanishing at $x_1\to \pm \infty$, of the form
\numparts
\begin{eqnarray}
\fl {\hat G} = \frac{s_{\nu}}{2 \sqrt{\kappa_0^2+l^2}}
\exp \left( - \sqrt{\kappa_0^2+l^2} \vert x_1-x_2 \vert \right)
\nonumber \\ + A_l \exp\left[ - \sqrt{\kappa_0^2+l^2} (x_1+x_2) \right],
\qquad & (x_1\in X_+), \label{3.36a} \\
\fl {\hat G} = \exp \left( - \sqrt{\kappa_0^2+l^2} x_2 \right)
\left[ B_l \exp \left( \sqrt{\kappa^2+l^2} x_1 \right) \right.
\nonumber \\ \left. + C_l \exp\left( - \sqrt{\kappa^2+l^2} x_1 \right) 
\right] , & (x_1\in X_0), \label{3.36b} \\
\fl {\hat G}  =  D_l \exp \left[ \sqrt{\kappa_0^2+l^2} (x_1 - x_2) \right], 
& (x_1\in X_-) . \label{3.36c}
\end{eqnarray}
\endnumparts
The constants $A_l, B_l, C_l$ and $D_l$ are determined by the conditions
that ${\hat G}$ and $\partial {\hat G}/\partial x_1$ are continuous at
$x_1=\pm d/2$.
Introducing the auxiliary function
\begin{eqnarray}
\fl W(l) = \left[ (\kappa_0^2+l^2) +(\kappa^2+l^2) \right]
\sinh \left( \sqrt{\kappa^2+l^2} d \right) \nonumber \\
+ 2 \sqrt{\kappa_0^2+l^2} \sqrt{\kappa^2+l^2}
\cosh \left( \sqrt{\kappa^2+l^2} d \right) , \label{3.37} 
\end{eqnarray}
one finds
\numparts
\begin{eqnarray}
\fl A_l = \frac{s_{\nu}}{2 W(l)} 
\frac{\kappa_0^2-\kappa^2}{\sqrt{\kappa_0^2+l^2}}
\exp\left( \sqrt{\kappa_0^2+l^2}d \right) 
\sinh\left( \sqrt{\kappa^2+l^2}d \right) , \label{3.38a} \\
\fl B_l = \frac{s_{\nu}}{2 W(l)} 
\left( \sqrt{\kappa_0^2+l^2}+\sqrt{\kappa^2+l^2} \right)
\exp\left[ \left( \sqrt{\kappa_0^2+l^2}+\sqrt{\kappa^2+l^2}
\right) \frac{d}{2} \right] , \label{3.38b} \\
\fl C_l  =  - \frac{s_{\nu}}{2 W(l)} 
\left( \sqrt{\kappa_0^2+l^2}-\sqrt{\kappa^2+l^2} \right)
\exp\left[ \left( \sqrt{\kappa_0^2+l^2}-\sqrt{\kappa^2+l^2}
\right) \frac{d}{2} \right] . \label{3.38c} 
\end{eqnarray}
\endnumparts

Let us denote by ${\bf f}_+$ the total average force per unit area
acting on the slab $\Lambda_+$ as a whole.
This force, directed along the $x$-axis perpendicular to the plates,
is determined by the contact particle densities as follows.
The electrolyte localized in the domain $\Lambda_0$ contributes
to this force by $\beta^{-1} n(d^-/2) {\bf e}_x$, where ${\bf e}_x$
is the unit vector along the $x$ axis.
The Coulomb gas localized in the considered slab $\Lambda_+$
contributes by $\beta^{-1} [ -n_0(d^+/2) + n_0^{\rm (vac)}] {\bf e}_x$,
where $n_0^{\rm (vac)}$ is the particle density at $x=d/2+L$, i.e.
the plain-hard-wall contact density for the Coulomb gas with the
bulk density equal to $n_0$.
The resulting force is given by
\begin{equation} \label{3.39}
\beta {\bf f}_+ = \left\{ n\left( \frac{d^-}{2} \right) -
\left[ n_0 \left( \frac{d^+}{2} \right) - n_0^{\rm (vac)} \right]
\right\} {\bf e}_x .
\end{equation}
The total force per unit surface acting on the slab $\Lambda_-$,
${\bf f}_-$, is for symmetry reasons opposite to ${\bf f}_+$,
${\bf f}_- = - {\bf f}_+$.
The force ${\bf f}$ by which the slab $\Lambda_-$ acts on the slab
$\Lambda_+$ is determined as the difference of the force given
by (\ref{3.39}) and its asymptotic $d\to\infty$ separation value,
\begin{equation} \label{3.40}
{\bf f}(d) = {\bf f}_+(d) - {\bf f}_+(\infty) .
\end{equation}
Equivalently,
\begin{equation} \label{3.41}
\beta {\bf f} = ( \delta n - \delta n_0 ) {\bf e}_x ,
\end{equation}
where
\begin{equation} \label{3.42}
\delta n = n\left( \frac{d^-}{2}\right) -
n\left( \frac{d^-}{2}\right)\Bigg\vert_{d\to\infty} , \quad
\delta n_0 = n_0\left( \frac{d^+}{2}\right) -
n_0\left( \frac{d^+}{2}\right)\Bigg\vert_{d\to\infty} .
\end{equation}
After some algebra one gets
\begin{eqnarray}
\fl \beta {\bf f}(d) = - \frac{(\kappa_0^2-\kappa^2)^2}{2^{\nu-1}
\pi^{(\nu-1)/2} \Gamma\left( \frac{\nu-1}{2} \right)} 
\int_0^{\infty} \rmd l \frac{l^{\nu-2} \sqrt{\kappa^2+l^2}}{\left( 
\sqrt{\kappa_0^2+l^2} + \sqrt{\kappa^2+l^2} \right)^2} \nonumber \\
\times \frac{1}{W(l)} \exp \left( - \sqrt{\kappa^2+l^2} d \right) 
{\bf e}_x .
\label{3.43}
\end{eqnarray}
$W(l)$ is defined by (\ref{3.37}).
Note that the force between the slabs is always attractive.
The large-distance asymptotic of this force depends
on whether $\kappa=0$ or $\kappa>0$.

When there is vacuum between the slabs, i.e. $\kappa\equiv 0$,
rescaling $l$ into $l=\kappa_0 k$, the amplitude $f$ of 
the attractive force can be expressed as
\begin{eqnarray}
\fl \beta f = \frac{\kappa_0^{\nu}}{2^{\nu-1} \pi^{(\nu-1)/2}
\Gamma\left( \frac{\nu-1}{2} \right)}
\int_0^{\infty} \frac{\rmd k \, k^{\nu-1}}{\left( \sqrt{1+k^2}
+ k \right)^2} \exp \left( - k \kappa_0 d \right) \nonumber \\
\times \left\{ \left( \sqrt{1+k^2}+k \right)^2 \sinh(k\kappa_0 d)
+ 2 k \sqrt{1+k^2} \exp( -k \kappa_0 d ) \right\}^{-1} . \label{3.44}
\end{eqnarray}
In the large-distance limit $\kappa_0 d \to \infty$, one introduces
the new integration variable $k'\in \langle 0,\infty)$ via
$k = k'/(\kappa_0 d)$, and then expands the integrated function in
$1/(\kappa_0 d)$, with the result
\begin{equation} \label{3.45}
\beta f = \frac{1}{d^{\nu}} 
\frac{(\nu-1) \Gamma(\nu/2) \zeta(\nu)}{2^{\nu} \pi^{\nu/2}}
\left\{ 1 - \frac{2\nu}{\kappa_0 d} + 
O\left( \frac{1}{(\kappa_0 d)^2} \right) \right\} . 
\end{equation}
The first term on the rhs of equation (\ref{3.45}) corresponds to the
electrostatic Casimir force in the ideal-conductor limit of the walls,
the second term describes the leading (still long-ranged) 
non-ideality correction due to the finite correlation length 
$1/\kappa_0$ of the Coulomb gas forming the slabs.
In the short-distance limit $\kappa_0 d\to 0$, one finds 
from (\ref{3.44}) that
\begin{equation} \label{3.46}
\beta f  =  \frac{\kappa_0^{\nu}}{2^{\nu} \pi^{(\nu-1)/2}
\Gamma\left( \frac{\nu-1}{2} \right)} \int_0^{\infty} 
\frac{\rmd k \, k^{\nu-2}}{\sqrt{1+k^2}}
\frac{1}{\left( \sqrt{1+k^2} + k \right)^2} .
\end{equation}
Explicitly,
\begin{equation} \label{3.47}
\beta f = \cases{ \kappa_0^2/(8\pi), & $(\nu=2)$, \\
\kappa_0^3/(24\pi), & $(\nu=3)$.} 
\end{equation}
In the short-distance limit, the density of particles at
$x=d^+/2$ is in fact the bulk density $n_0$, so that
$\beta f = n_0 - n_0^{\rm (vac)}$.
According to \cite{Jancovici4},
$n_0^{\rm (vac)} = n_0 -\kappa_0^2/(8\pi)$ in 2D and
$n_0^{\rm (vac)} = n_0 -\kappa_0^3/(24\pi)$ in 3D,
which confirms the obtained results (\ref{3.47}).

When there is an electrolyte between the slabs, i.e. $\kappa>0$,
the amplitude $f$ of the attractive force between the slabs,
see equation (\ref{3.43}), can be expressed in the large-separation
limit $d \to \infty$ as follows
\begin{equation} \label{3.48}
\fl \beta f { \atop \stackrel{\sim}{\scriptstyle{d\to\infty}} }
\frac{2 (\kappa_0^2 - \kappa^2)^2}{(4\pi)^{(\nu-1)/2}
\Gamma\left( \frac{\nu-1}{2} \right)} 
\int_0^{\infty} \frac{ \rmd l \, l^{\nu-2} 
\sqrt{\kappa^2+l^2}}{\left( \sqrt{\kappa_0^2+l^2}
+ \sqrt{\kappa^2+l^2} \right)^4} 
\exp \left( - 2 \sqrt{\kappa^2+l^2} d \right) . 
\end{equation}
Introducing the new integration variable $k\in\langle 0,\infty)$ via
$\sqrt{\kappa^2+l^2} = \kappa +k/d$ and then taking the limit
$\kappa d\to\infty$, one finds the exponentially decaying force
\begin{equation} \label{3.49}
\beta f { \atop \stackrel{\sim}{\scriptstyle{d\to\infty}} } 
\frac{(\kappa_0-\kappa)^2}{(\kappa_0+\kappa)^2}
\frac{\kappa^{\nu}}{(4\pi\kappa d)^{(\nu-1)/2}}
\exp ( - 2 \kappa d ) .
\end{equation}
If the slabs are made from ideal conductors ($\kappa_0\to \infty$) or 
they are empty ($\kappa_0 = 0$, i.e. plain hard walls with $\epsilon =1$)
this short-distance force takes the form
\begin{equation} \label{3.50}
\beta f { \atop \stackrel{\sim}{\scriptstyle{d\to\infty}} } 
\frac{\kappa^{\nu}}{(4\pi\kappa d)^{(\nu-1)/2}}
\exp ( - 2 \kappa d ) .
\end{equation}
Note that the parameter $(2\kappa)$ in the exponential decay
corresponds to the inverse correlation length of the density-density
distribution functions in the Debye-H\"uckel limit \cite{Samaj5}.
We conclude that the presence of an electrolyte between the conducting
slabs induces screening of the long-ranged Casimir force.

\subsection{Plasma between dielectric walls}
We consider the case of a Coulomb gas between dielectric walls. 
We do not use any microscopic description of the walls and resort to the
phenomenological separation (\ref{2.9}). 
The derivative $\partial F_0/\partial d$ where $F_0$ is the free energy of 
the wall system alone has already been calculated in section 2.2. 
We now use the Debye-H\"uckel theory for computing the derivative 
$\partial\Omega_{\rm eff}/\partial d$ where $\Omega_{\rm eff}$ is the
grand potential of the Coulomb gas computed with the Hamiltonian 
$H_{\rm eff}$, i.e. with inert dielectric walls. 
The geometry is the same as in section 2.2, except that now a Coulomb gas 
with the Debye inverse length $\kappa$ fills the slab $-d/2 < x < d/2$.

We shall use (\ref{3.14}), thus we need the screened potential $G$. 
Let the source point ${\bf r}_2=(x_2,{\bf r}_2^{\perp})$ 
be in the Coulomb gas. 
The Fourier transform $\hat{G}$ of the screened potential is determined by 
the equations
\numparts
\begin{eqnarray}
\fl \left(\frac{\partial^2}{\partial x_1^2}-\kappa^2-l^2\right)
{\hat G}(x_1,x_2;{\bf l}) = -s_{\nu}\delta(x_1-x_2),
\qquad  & \left( -d/2 < x_1 < d/2 \right) , \label{3.51a}   \\
\fl \left(\frac{\partial^2}{\partial x_1^2}-l^2\right)
{\hat G}(x_1,x_2;{\bf l}) =  0,
& \mbox{($x_1<-d/2$ or $x_1>d/2$),} \label{3.51b}  
\end{eqnarray}
\endnumparts
with the conditions that ${\hat G}$ and 
$\epsilon (\partial{\hat G}/\partial x_1)$ (with $\epsilon=1$ in the
plasma region) be continuous at $x_1=\pm d/2$ and that 
${\hat G}\rightarrow 0$ when $x_1\rightarrow\pm\infty$. 
A derivation similar to the one in the previous section 3.3 gives, 
when both $x_1$ and $x_2$ are in the Coulomb gas,
\begin{eqnarray}
\fl {\hat G}(x_1,x_2;{\bf l}) = \frac{s_{\nu}}{\sqrt{\kappa^2+l^2}}
\Bigg\{\frac{1}{2}\exp\left(-\sqrt{\kappa^2+l^2}|x_1-x_2| \right)
\nonumber \\
\fl + \frac{b_l \cosh\left[ \sqrt{\kappa^2+l^2}(x_1+x_2)\right]
+ b_l^2\exp\left( -\sqrt{\kappa^2+l^2} d \right)
\cosh \left[ \sqrt{\kappa^2+l^2}(x_1-x_2) \right] }
{ \exp\left( \sqrt{\kappa^2+l^2} d\right)
-b_l^2\exp\left( - \sqrt{\kappa^2+l^2} d\right) } \Bigg\}
\label{3.52}
\end{eqnarray}
where
\begin{equation} \label{3.53}
b_l = \frac{\sqrt{\kappa^2+l^2} - \epsilon l}
{\sqrt{\kappa^2+l^2} + \epsilon l}.
\end{equation}

We now use (\ref{3.52}) in (\ref{3.14}). 
The first term in the curly brackets of (\ref{3.52}) corresponds 
to the bulk $G$; using the bulk relation (\ref{3.5}) and linearizing 
the exponential in (\ref{3.14}) gives
\begin{equation} \label{3.54}
\fl n(x) - n = -\frac{\beta n}{2}\int\frac{\rmd^{\nu -1}l}
{(2\pi)^{\nu -1}} \frac{s_{\nu}}{\sqrt{\kappa^2+l^2}}
\frac{b_l \cosh\left( 2\sqrt{\kappa^2+l^2} x \right)
+ b_l^2\exp\left(-\sqrt{\kappa^2+l^2} d \right)}
{\exp\left(\sqrt{\kappa^2+l^2}d\right)
- b_l^2\exp\left(-\sqrt{\kappa^2+l^2}d \right)}.
\end{equation} 
Therefore
\begin{eqnarray} \label{3.55}
\fl \int_{-d/2}^{d/2}\rmd x [n(x)-n] = - \frac{\beta n}{2}
\int\frac{\rmd^{\nu -1}l}{(2\pi)^{\nu -1}}
\frac{s_{\nu}}{\kappa^2+l^2} \nonumber \\
\times \frac{b_l \sinh\left( \sqrt{\kappa^2+l^2}d \right)
+ b_l^2 \sqrt{\kappa^2+l^2} d \exp\left( -\sqrt{\kappa^2+l^2}d \right)}
{\exp\left(\sqrt{\kappa^2+l^2} d \right)
-b_l^2\exp\left(-\sqrt{\kappa^2+l^2} d \right)}.
\end{eqnarray}

The grand potential $\Omega_{\rm eff}$ is the sum of a bulk part and a
surface part (due to the walls) $\Omega_W$. 
This surface part is related to the surface part 
of the number of particles (\ref{3.55}) by
\begin{equation} \label{3.56} 
n\frac{\partial (-\beta\Omega_W)}{\partial n}
={\mathcal A}\int_{-d/2}^{d/2}\rmd x [n(x)-n]
\end{equation}  
since, at the lowest order in $\beta$, we can replace 
$z\partial /\partial z$ by $n\partial /\partial n$. 

In the limit $d\rightarrow\infty$, $\Omega_W/{\mathcal A}$ reduces to
twice the surface tension $\gamma$, and (\ref{3.56}) with the limiting
form of (\ref{3.55}) give
\begin{equation} \label{3.57}
n\frac{\partial (\beta\gamma)}{\partial n}=\frac{\kappa^2}{8}
\int\frac{\rmd^{\nu -1}l}{(2\pi)^{\nu -1}}
\frac{b_l}{\kappa^2+l^2}.
\end{equation}
The rescaling of $l$ into $l=\kappa k$ makes apparent that the r.h.s. of
(\ref{3.57}) depends on $n$ by a factor $\kappa^{\nu -1}$. 
Using $n\partial /\partial n=(1/2)\kappa\partial /\partial\kappa$, 
we can integrate (\ref{3.57}) into
\begin{equation} \label{3.58}
\beta \gamma = \frac{\kappa^{\nu -1}}{4(\nu -1)}
\int\frac{\rmd^{\nu -1}k}{(2\pi)^{\nu -1}}
\frac{1}{1+k^2}\frac{\sqrt{1+k^2}-\epsilon k}
{\sqrt{1+k^2}+\epsilon k}.
\end{equation}

In 2D, performing the integral in (\ref{3.58}) gives
\begin{equation} \label{3.59}
\beta\gamma = \cases{
\frac{\kappa}{2\pi}\left[-\frac{\pi}{4}
+\frac{2}{\sqrt{1-\epsilon^2}}\tan^{-1}\left(\frac{1-\epsilon}
{1+\epsilon}\right)^{1/2}\right] & if $\epsilon<1$, \\
\frac{\kappa}{2\pi}\left[-\frac{\pi}{4}
+\frac{2}{\sqrt{\epsilon^2-1}}\tanh^{-1}\left(\frac{\epsilon -1}
{\epsilon +1}\right)^{1/2}\right] & if $\epsilon>1$.}
\end{equation}

In 3D, for $\epsilon=1$, (\ref{3.58}) reproduces the $\nu =3$ surface
tension (\ref{3.27}). 
For other values of $\epsilon$, the integral in (\ref{3.58}) diverges. 
When $\epsilon>1$, this divergence is to be expected for point particles, 
since the interaction between a particle and its image is attractive 
and produces a non-integrable Boltzmann factor. 
However, when  $\epsilon<1$, the divergence comes from the 
inappropriate linearization of the exponential in (\ref{3.14}); a more
careful treatment \cite{Onsager,Dean} gives a finite surface tension.

For a finite distance $d$, the $d$-dependent part of the grand potential
per unit area is 
$\omega_{\rm eff}(d) = (\Omega_W/{\mathcal A})-2\gamma$. 
After the substraction of the infinite-$d$ part of (\ref{3.55}), using again 
$n\partial /\partial n=(1/2)\kappa\partial /\partial\kappa$, we now obtain
\begin{eqnarray} 
\fl \frac{\partial(\beta\omega_{\rm eff})}{\partial\kappa} =
\frac{1}{2(2\pi)^{\nu-1}} \int{\rm d}^{\nu-1}l \nonumber \\
\times \frac{\kappa[b_l(b_l^2-1)(\kappa^2+l^2)^{-1}
+2b_l^2d(\kappa^2+l^2)^{-1/2}]\exp(-2d\sqrt{\kappa^2+l^2})}
{1-b_l^2\exp(-2d\sqrt{\kappa^2+l^2})}. \label{3.60}     
\end{eqnarray}
It can be checked that the numerator in the integrand of (\ref{3.60}) is
the derivative with respect to $\kappa$ of the denominator. 
Thus, integrating (\ref{3.60}) with the condition $\omega_{\rm eff}=0$ 
when $\kappa=0$ gives
\begin{eqnarray} 
\fl \beta\omega_{\rm eff} = \frac{1}{2(2\pi)^{\nu-1}}\int{\rm d}^{\nu-1}l
\Bigg\{\ln\left[1-b_l^2\exp(-2d\sqrt{\kappa^2+l^2})\right] \nonumber \\ 
- \ln\left[1-\left(\frac{1-\epsilon}{1+\epsilon}\right)^2\exp(-2dl)\right]
\Bigg\}. \label{3.61}                               
\end{eqnarray}

The second logarithm in equation (\ref{3.61}) gives to the force 
$-\partial\omega_{\rm eff}/\partial d$ a long-range contribution
opposite to the Casimir force in vacuum (\ref{2.15}): the total
long-range force associated to 
$\omega = \omega_{\rm eff}+(F_0/{\mathcal A})$ vanishes. 
The cancellation (screening) effect does occur in the present model.
Only a short-range $d$-dependent force remains, associated to 
\begin{equation} \label{3.62}
\beta\omega=\frac{1}{2(2\pi)^{\nu-1}}\int{\rm d}^{\nu-1}l\,
\ln\left[1-b_l^2\exp(-2d\sqrt{\kappa^2+l^2}\,)\right]. 
\end{equation}
Thus, in the 3D case $\nu=3$, we have retrieved by our method a
previously known result \cite{Mahanty,Dean}.

In the limit $\kappa d\rightarrow\infty$, the asymptotic behaviour of 
(\ref{3.62}) is
\begin{equation} \label{3.63}
\beta\omega\sim -\frac{1}{2(2\pi)^{\nu-1}}\int{\rm d}^{\nu-1}l\,
b_l^2\exp(-2d\sqrt{\kappa^2+l^2}).   
\end{equation}
In this equation, only small values of $l/\kappa$ contribute to the 
integral. 
Thence, $\exp(-2d\sqrt{\kappa^2+l^2})\sim\exp(-2\kappa d-dl^2/\kappa)$ 
and $b_l^2\sim 1$. 
With these replacements in (\ref{3.63}),
\begin{equation} \label{3.64} 
\beta\omega\sim -\frac{\kappa^{(\nu -1)/2}\exp(-2\kappa d)}
{2^{\nu}(\pi d)^{(\nu -1)/2}},   
\end{equation}
and the force per unit area has the asymptotic behaviour for large $\kappa d$
\begin{equation} \label{3.65}
-\frac{\partial\omega}{\partial d}\sim-\frac{\beta^{-1}\kappa^{\nu}
\exp(-2\kappa d)}{(4\pi\kappa d)^{(\nu -1)/2}}.   
\end{equation}
This asymptotic behaviour has the remarkable feature that it does not
depend on $\epsilon$. 
This was already apparent on (\ref{3.49}) in the special cases 
$\epsilon=1\;(\kappa_0=0)$ and 
$\epsilon\rightarrow\infty\;(\kappa_0\rightarrow\infty)$.

\section{The free-fermion point}

\subsection{General formalism}
The 2D symmetric Coulomb gas is exactly solvable at the collapse coupling
$\beta=2$ by mapping onto the Thirring model at the free fermion point.
We first shortly review the bulk regime.
Both species of particles have the same rescaled fugacity $m = 2\pi a z$ 
[where $a$ is the length scale introduced in equation (\ref{1.2})]
which has the dimension of an inverse length.
The general formalism \cite{Cornu1,Cornu2} expresses the many-particle
densities in terms of specific Green functions $G_{qq'}({\bf r},{\bf r}')$
$(q,q'=\pm)$.
Because of the symmetry between positive and negative particles, one only
needs $G_{++}$ and $G_{-+}$ which are determined by
\begin{equation} \label{4.1}
( \Delta_1 - m^2 ) G_{++}({\bf r}_1,{\bf r}_2) = 
- m \delta({\bf r}_1-{\bf r}_2)
\end{equation}
and 
\begin{equation} \label{4.2}
G_{-+}({\bf r}_1,{\bf r}_2) =  - \frac{1}{m}
\left( \frac{\partial}{\partial x_1} + 
\rmi \frac{\partial}{\partial y_1} \right)
G_{++}({\bf r}_1,{\bf r}_2) .
\end{equation}
These equations have to be supplemented by the vanishing boundary conditions
when $\vert {\bf r}_1-{\bf r}_2 \vert \to \infty$.
In infinite space, the solution of equations (\ref{4.1}) and (\ref{4.2})
reads
\numparts
\begin{eqnarray}
G_{++}({\bf r}_1,{\bf r}_2) = \frac{m}{2\pi}
K_0(m\vert {\bf r}_1-{\bf r}_2 \vert) , \label{4.3a} \\
G_{-+}({\bf r}_1,{\bf r}_2) = \frac{m}{2\pi}
\frac{(x_1-x_2)+ \rmi (y_1-y_2)}{\vert {\bf r}_1-{\bf r}_2 \vert}
K_1(m\vert {\bf r}_1-{\bf r}_2 \vert) , \label{4.3b}
\end{eqnarray}
\endnumparts
where $K_0$ and $K_1$ are modified Bessel functions.
The one-particle densities, given by
\begin{equation} \label{4.4}
n_q = m G_{qq}({\bf r},{\bf r}),
\end{equation}
are infinite since $K_0(m r)$ diverges logarithmically as $r\to 0$.
This divergence can be suppressed by a short-distance cutoff $R$.
We replace the point-like particles by small charged hard disks
of diameter $R$ and use a regularized form of (\ref{4.4})
for the total particle density $n = n_+ + n_-$,
\begin{equation} \label{4.5}
n = \frac{m^2}{\pi} K_0(m R)
{ \atop \stackrel{\sim}{\scriptstyle{m R \to 0}} }
\frac{m^2}{\pi} \left[ {\rm ln}\left( \frac{2}{m R} \right) - C \right] . 
\end{equation}  
Such a regularization is consistent in the sense that 
the perfect-screening sum rule is satisfied and the expected
equation of state is reproduced \cite{Cornu1,Cornu2}.

The above formalism can be generalized to inhomogeneous situations
when the system domain $\Lambda = \cup_{\alpha} \Lambda^{(\alpha)}$.
Each subdomain $\Lambda^{(\alpha)}$ is characterized by a constant
rescaled fugacity, $m({\bf r}) = m_{\alpha}$ for 
${\bf r}\in \Lambda^{(\alpha)}$, and the corresponding bulk density
$n_{\alpha}$ defined as a function of $m_{\alpha}$ by (\ref{4.5}).
The Green function $G_{++}$ obeys the differential equation
\begin{equation} \label{4.6}
( \Delta_1 - m_{\alpha}^2 ) G_{++}({\bf r}_1,{\bf r}_2)
= - m_{\alpha} \delta({\bf r}_1 - {\bf r}_2), \quad \quad
{\bf r}_1 \in \Lambda^{(\alpha)},
\end{equation}
where the spatial position of the source point ${\bf r}_2$ is
arbitrary.
The Green function $G_{-+}$ is determined by $G_{++}$ via 
equation (\ref{4.2}) with $m$ taken as the subdomain-dependent
$m({\bf r}_1)$.
The boundary conditions are that $G_{++}$ and $G_{-+}$ must be
continuous at each subdomain boundary $\partial \Lambda^{(\alpha)}$.
The one-particle densities are again given by (\ref{4.4}).

\subsection{Polarizable interface}
In the geometry of figure 1, the rescaled particle fugacity is equal
to $m$ in the half-space $x>0$ and to $m_0$ in the half-space $x<0$.
Let the source point ${\bf r}_2 = (x_2,y_2)$ has the coordinate $x_2>0$.
In the Fourier representation (\ref{3.15}), equation (\ref{4.6}) 
takes the form
\numparts
\begin{eqnarray}
\left( \frac{\partial^2}{\partial x_1^2} - m^2 - l^2 \right)
{\hat G}_{++}(x_1,x_2;l) =  - m \delta(x_1-x_2), \qquad
& ( x_1 > 0 ), \label{4.7a} \\
\left( \frac{\partial^2}{\partial x_1^2} - m_0^2 - l^2 \right)
{\hat G}_{++}(x_1,x_2;l) =  0, & ( x_1 < 0 ). \label{4.7b}
\end{eqnarray}
\endnumparts
We look for a solution, vanishing at $x_1\to\pm\infty$, of the form
\numparts
\begin{eqnarray}
\fl {\hat G}_{++} = \frac{m}{2\sqrt{m^2+l^2}} 
\exp \left( - \sqrt{m^2+l^2}\vert x_1-x_2 \vert \right) \nonumber \\
+ A_l \exp \left[ - \sqrt{m^2+l^2} (x_1+x_2) \right] ,
& (x_1>0) , \label{4.8a} \\
\fl {\hat G}_{++} =  B_l 
\exp \left( \sqrt{m_0^2+l^2} x_1 - \sqrt{m^2+l^2} x_2 \right) , \qquad 
& (x_1<0) . \label{4.8b}
\end{eqnarray}
\endnumparts
The constants $A_l$ and $B_l$ are determined by the conditions that
${\hat G}_{++}$ and 
${\hat G}_{-+} = [m(x_1)]^{-1} (l-\partial_{x_1}) {\hat G}_{++}$ 
are continuous at $x_1 = 0$.
The final result is
\numparts
\begin{eqnarray}
A_l = - \frac{m}{2 \sqrt{m^2+l^2}} + B_l , \label{4.9a} \\
B_l = \left( \frac{\sqrt{m^2+l^2}+l}{m} +
\frac{\sqrt{m_0^2+l^2}-l}{m_0} \right)^{-1} . \label{4.9b}
\end{eqnarray}
\endnumparts
The particle density at $x>0$ is then given by
\begin{equation} \label{4.10}
n(x) = n + \frac{m}{\pi} \int_0^{\infty} \rmd l
\left( A_l + A_{-l} \right) 
\exp \left( - 2 \sqrt{m^2+l^2} x \right) ,
\end{equation}
where $n$ is the regularized bulk density related to the rescaled
fugacity $m$ via (\ref{4.5}).
The same procedure applies when the source point 
${\bf r}_2 = (x_2,y_2)$ has the component $x_2<0$.
For symmetry reasons, one gets the particle density at $x<0$
in the form $n_0(x) = n_0 +$the last term of equation (\ref{4.10})
with the interchange $m\leftrightarrow m_0$. 

The surface tension $\gamma$ is again determined by the relations
\numparts
\begin{eqnarray}
m \frac{\partial (-\beta\gamma)}{\partial m}
= \int_0^{\infty} \rmd x \left[ n(x)-n \right] , \label{4.11a} \\
m_0 \frac{\partial (-\beta\gamma)}{\partial m_0}
= \int^0_{-\infty} \rmd x \left[ n_0(x)-n_0 \right] , \label{4.11b}
\end{eqnarray}
\endnumparts
with the boundary condition $\beta\gamma(m,m_0) = 0$ when $m=m_0$.
The result is
\begin{equation} \label{4.12}
\beta \gamma = \int_0^{\infty} \frac{\rmd l}{2\pi} \,
{\rm ln} \left( \frac{2 \sqrt{m^2+l^2} \sqrt{m_0^2+l^2}}{
m m_0 + \sqrt{m^2+l^2} \sqrt{m_0^2+l^2} + l^2} \right) .
\end{equation}
When there are no particles in the half-space $x<0$, i.e. $m_0=0$, 
(\ref{4.12}) yields the surface tension of the Coulomb gas
in contact with a vacuum,
\begin{equation} \label{4.13}
\beta \gamma^{\rm (vac)}(m) = \frac{m}{4\pi} (\pi - 2) ,
\end{equation}
in full agreement with the result obtained in \cite{Cornu1}.

In the ideal-conductor limit of the Coulomb gas in the half-space
$x<0$, the surface tension of equation (\ref{4.12}) has the following
large-$m_0$ expansion
\begin{equation} \label{4.14}
\beta \gamma = \frac{m_0}{4\pi} (\pi - 2)
+ m \left[ \frac{1}{2\pi} {\rm ln} \left( \frac{m}{m_0} \right)
+ O(1) \right] .
\end{equation}
The first term on the rhs of equation (\ref{4.14}) is nothing but the
vacuum surface tension [see (\ref{4.13})] of the Coulomb gas
in the half-space $x<0$.
The second term, the surface tension of the Coulomb gas at $x>0$
induced by the ideal-conductor wall at $x<0$, diverges due to the
particle-image collapse at $\beta=2$ \cite{Samaj3}.

\subsection{Two-slab geometry}
The above method can be readily extended to the two-slab geometry 
of figure 2.
We keep the notation of intervals $X_+ = \langle d/2,\infty)$,
$X_0 = \langle -d/2,d/2 \rangle$ and $X_- = (-\infty,-d/2\rangle$
from section 3.3.
If the source point ${\bf r}_2 = (x_2,y_2)$ has the coordinate
$x_2\in X_+$, the solution of equations (\ref{4.6}) is searched, 
in the Fourier space, in the form
\numparts
\begin{eqnarray}
\fl {\hat G}_{++} = \frac{m_0}{2 \sqrt{m_0^2+l^2}}
\exp\left( - \sqrt{m_0^2+l^2} \vert x_1-x_2 \vert \right) \nonumber \\
+ A_l \exp \left[ - \sqrt{m_0^2+l^2} (x_1+x_2) \right] , \qquad 
& ( x_1 \in X_+ ) , \label{4.15a} \\
\fl {\hat G}_{++} = \exp \left( - \sqrt{m_0^2+l^2} x_2 \right)
\left[ B_l \exp\left( \sqrt{m^2+l^2} x_1 \right) \right. \nonumber \\
\left. + C_l \exp\left( - \sqrt{m^2+l^2} x_1 \right) \right] ,
& ( x_1 \in X_0 ) , \label{4.15b} \\
\fl {\hat G}_{++} = D_l \exp\left[ \sqrt{m_0^2+l^2} (x_1-x_2) \right] ,
& ( x_1 \in X_- ) . \label{4.15c} 
\end{eqnarray}
\endnumparts
The unknown constants $A_l$, $B_l$, $C_l$ and $D_l$ are 
determined by the conditions that ${\hat G}_{++}$ and 
$[m(x_1)]^{-1} (l-\partial_{x_1}) {\hat G}_{++}$ 
are continuous at $x_1 = \pm d/2$.
The particle density at $x\ge d/2$ is determined by
\begin{equation} \label{4.16}
n_0(x) = n_0 + \frac{m_0}{\pi} \int_0^{\infty} \rmd l
\left( A_l + A_{-l} \right) 
\exp\left( - 2 \sqrt{m_0^2+l^2} x \right) .
\end{equation}
The same formalism applies when the source point ${\bf r}_2 = (x_2,y_2)$
has the coordinate $x_2\in X_0$, to obtain the density profile
between the slabs.

The force between the slabs is calculated by using the procedure
outlined in section 3.3, formulae (\ref{3.39}) - (\ref{3.43}).
Without going into technical details, we have found that the force 
${\bf f}(d)$, defined by (3.40), is always attractive.
Its amplitude $f$ as a function of the distance $d$ between the slabs
is given by
\numparts
\begin{eqnarray}
\fl \beta f(d) = \frac{(m_0-m)^2}{\pi} \int_0^{\infty} \rmd l
\frac{l^2 \sqrt{m^2+l^2} }{m m_0 + \sqrt{m^2+l^2} \sqrt{m_0^2+l^2}+l^2}
\nonumber \\
\times \frac{1}{W(l)} \exp\left( - \sqrt{m^2+l^2} d \right) ,
\label{4.17a}
\end{eqnarray}
where
\begin{eqnarray}
\fl W(l) = \left( m m_0 + \sqrt{m^2+l^2} \sqrt{m_0^2+l^2} + l^2 \right)
\sinh \left( \sqrt{m^2+l^2} d \right) \nonumber \\
+ \sqrt{m^2+l^2} \sqrt{m_0^2+l^2} \exp \left( - \sqrt{m^2+l^2} d \right) . 
\label{4.17b}
\end{eqnarray}
\endnumparts
The large-distance behaviour of $f(d)$ depends on whether the rescaled
fugacity $m=0$ or $m>0$.

When there is vacuum between the slabs, i.e. $m\equiv 0$,
the amplitude of the attractive force can be expressed in the limit
$m_0 d \to \infty$ as follows
\begin{equation} \label{4.18}
\beta f = \frac{\pi}{24 d^2} \left\{ 1 - \frac{4}{(2 m_0) d}
+ O \left( \frac{1}{(m_0 d)^2} \right) \right\} .
\end{equation}
The first term on the rhs of equation (\ref{4.18}) corresponds to the
2D electrostatic Casimir force in the ideal-conductor limit of the
slabs, the second term corresponds to the correction due to the finite
correlation length $1/(2 m_0)$ of the Coulomb gas forming the slabs.
It is interesting that this correction, evaluated at the $\beta=2$
collapse point, has the same form (in terms of the correlation length)
as the one evaluated in the Debye-H\"uckel $\beta\to 0$ limit,
see formula (\ref{3.45}).

When there is an electrolyte between the slabs, i.e. $m>0$,
the amplitude of the attractive force has the following large-$d$
behaviour
\begin{equation} \label{4.19}
\beta f { \atop \stackrel{\sim}{\scriptstyle{d\to\infty}} } 
\cases{
\frac{(m_0-m)^2}{m_0^2} \frac{m^2}{8\sqrt{\pi}(m d)^{3/2}}
\exp(-2 m d) & for $m_0>0$, \\
\frac{m^2}{\sqrt{\pi m d}} \exp(-2 m d) & for $m_0=0$ .}
\end{equation}
If the slabs are made from an ideal-conductor material with
$m_0\to\infty$, this short-distance force is written as
\begin{equation} \label{4.20}
\beta f { \atop \stackrel{\sim}{\scriptstyle{d\to\infty}} } 
\frac{m^2}{8\sqrt{\pi}(m d)^{3/2}} \exp(-2 m d) .
\end{equation}
These formulae confirm that, also at the special $\beta=2$ coupling,
an electrolyte between the ideal-conductor slabs causes screening
of the long-range Casimir force.
Comparing with the Debye-H\"uckel result [see equations (\ref{3.49}) 
and (\ref{3.50})] one sees that the residual short-range 
force is temperature-dependent.
Note that at $\beta=2$ the residual force depends on the dielectric 
constant $\epsilon$ of the slabs, since the results (\ref{4.19}) for
$\epsilon = 1$ ($m_0 = 0$) and (\ref{4.20}) for $\epsilon\to\infty$ 
are different.
The non-dependence on $\epsilon$ found in (\ref{3.50}) and (\ref{3.65}) 
seems to hold only in the Debye-H\"uckel regime.

\section{Conclusion} 
On fully microscopic models, in the classical (i.e. non quantum) regime,
we have studied the properties of a Coulomb gas near a plane conducting
wall, and two parallel conducting walls with or without a Coulomb gas
between them. We have also considered the case of dielectric walls of
arbitrary dielectric constant $\epsilon$, without a microscopic model
for these walls, using macroscopic electrostatics and linear response
theory. We have checked that the long-range Casimir force between
parallel walls separated by a vacuum is screened into some (still
attractive) residual short-range force when a Coulomb gas is present
between the walls.

All these calculations have been performed taking into account only the
electrostatic part of the interactions. How to take into account the
magnetic field for dealing, on a microscopic model in the classical
limit, with the full particle-radiation interaction is an open problem.

\ack
The authors acknowledge support from the CNRS-SAS agreement,
Project No. 14439.
A partial support of L. \v Samaj by a VEGA grant is acknowledged.

\end{document}